\documentclass[oribibl]{llncs}

\usepackage{amssymb}	 % For \union.
\usepackage{amsmath}
\usepackage{balance}
\usepackage{graphicx}
\usepackage{booktabs}
\usepackage{multirow}
\usepackage{tabularx}
\usepackage{threeparttable}
\usepackage{url}

\newcommand{\mtw}{\textup{\textbf{\textrm{W}}}}
\newcommand{\mtc}{\textup{\textbf{\textrm{C}}}}
\newcommand{\mts}{\textup{\textbf{\textrm{S}}}}
\newcommand{\mtu}{\textup{\textbf{\textrm{U}}}}
\newcommand{\mtv}{\textup{\textbf{\textrm{V}}}}
\DeclareMathOperator{\TF}{TF}
\DeclareMathOperator{\TFLOG}{TF-LOG}
\DeclareMathOperator{\TFBOOL}{TF-BOOLEAN}
\DeclareMathOperator{\TFIDF}{TF-IDF}
\DeclareMathOperator{\DLMIDF}{DLM-IDF}

\usepackage{fancyhdr} % for arxiv
\fancypagestyle{firststyle}
{
   \fancyhf{}
   \fancyfoot[C]{\scriptsize{Proceedings of the 29th International Conference on Database and Expert Systems Applications (DEXA 2018), Regensburg, Springer, pp.~265--277. The final authenticated publication is available online at \url{https://doi.org/10.1007/978-3-319-99133-7_22}}}
}

\begin{document}

\title{Toward Validation of Textual Information \\ Retrieval Techniques for Software Weaknesses}
\author{Jukka Ruohonen%\orcidID{0000-0001-5147-3084} 
\and Ville Lepp\"anen \\ \email{\{juanruo, ville.leppanen\}@utu.fi}}
\institute{Department of Future Technologies, University of Turku, Finland}

\maketitle

\begin{abstract}
This paper presents a preliminary validation of common textual information retrieval techniques for mapping unstructured software vulnerability information to distinct software weaknesses. The validation is carried out with a dataset compiled from four software repositories tracked in the Snyk vulnerability database. According to the results, the information retrieval techniques used perform unsatisfactorily compared to regular expression searches. Although the results vary from a repository to another, the preliminary validation presented indicates that explicit referencing of vulnerability and weakness identifiers is preferable for concrete vulnerability tracking. Such referencing allows the use of keyword-based searches, which currently seem to yield more consistent results compared to information retrieval techniques. Further validation work is required for improving the precision of the techniques, however.
\end{abstract}

\begin{keywords}
text mining, software vulnerability, Snyk, LSA, CVE, CWE, NVD
\end{keywords}

\section{Introduction}

\thispagestyle{firststyle} % for arxiv

Software weaknesses---as cataloged in the so-called Common Weakness Enumeration (CWE) framework---are abstractions for security-related mistakes made in software development. Such weaknesses may lead to concrete software vulnerabilities. The CWE framework covers numerous different weaknesses, ranging from software design flaws to inadequate input validation, insecure maintaining of time and state, and lack of encapsulation \cite{Tsipenyuk05}. The richness of the framework has ensured its usefulness for both research and practice. To name a few of the application domains, the CWE framework has been used for security (compliance) assessments~\cite{GosevaPopstojanova17, Hale17}, risk analysis~\cite{Alsaleh17, Franqueira11}, quantitative trend analysis~\cite{Murtaza16}, data mining \cite{Jiminez16}, static source code analysis \cite{Oyetoyan18}, dissemination of fuzzing results \cite{KangPark17}, and last but not least, education and security awareness \cite{MartinBarnum08}. Also text mining applications have been common, although there are still gaps in the literature.

Many of the text mining applications have relied on the  Open Web Application Security Project (OWASP), which limits the generalizability of the applications~\text{\cite{RomanMunoz18, Peclat18}}. Another gap relates to the common focus on CWE-based ontologies. Although such ontologies are useful for understanding and clustering weaknesses~\text{\cite{Bojanova16, HanLi18, WuGandhi10}}, these have a limited appeal for vulnerability tracking. Here, the term vulnerability tracking refers to the concrete, largely manual software engineering work required for archiving and documenting software vulnerabilities. If this work does not explicitly cover weaknesses, an application of an ontology-based technique must first solve the problem of extracting the CWEs from the typically more or less unstructured textual vulnerability data. Limited attention has also been given for the validity of the text mining applications.

To contribute toward sealing some of these gaps, this paper tentatively examines the validity of common textual information retrieval techniques for extracting CWEs from vulnerability databases. The extraction itself has practical value because many vulnerability databases do not catalog weaknesses, partially due to the complexity of the CWE framework and the manual work required~\cite{WenZhang15}. By superseding the manual assignment of security bug reports to CWEs~\cite{GosevaPopstojanova17}, automatic extraction can also facilitate empirical security research. 
It is important to further emphasize that the task differs from conventional information retrieval systems that can return multiple documents for a single query. In contrast, this paper adopts a much stricter constraint: each unique vulnerability should map to a single unique CWE-identified weakness. As elaborated in the opening Section~\ref{section: materials}, this constraint can be also used for comparing common textual information retrieval techniques against simple regular expression searches. The comparative results are presented in Section~\ref{section: results} and discussed in Section~\ref{section: discussion}.

\section{Materials}\label{section: materials}

The following will outline the data sources, the subset of data used for the validation, the pre-processing routines, and the weights used for computation.

\subsection{Data Sources}\label{subsec: data sources}

The dataset is compiled from three distinct but related sources. The first source is the conventional National Vulnerability Database (NVD) maintained by the National Institute of Standards and Technology (NIST) in cooperation with the non-profit MITRE corporation. This database provides one-to-one mappings between abstract weaknesses identified with CWEs and concrete vulnerabilities identified with Common Vulnerabilities and Exposures (CVEs). These mappings are based on expert opinion; during the archival of vulnerabilities to the database, NVD's maintainers derive the weaknesses from the concrete vulnerabilities archived. At the time of retrieving the database's content~\cite{NVD18b}, there were $102$ unique CWEs in the database once rejected CVEs were excluded. (These invalid cases are marked with the string \texttt{REJECTED} in the summary field of a CVE.) These CWEs were used for assembling the estimation subset soon discussed.

The second source is the CWE database maintained by MITRE in cooperation with volunteers and governmental sponsors. In total, there were $730$ documented weaknesses in the database at the time of retrieval~\cite{MITRE18c}. These weaknesses have been used to construct different ontologies~\cite{HanLi18, Peclat18, WuGandhi10}, including the famous ``seven pernicious kingdoms'' of security-related programming mistakes~\cite{Tsipenyuk05}. Reflecting such ontologies, MITRE provides also many predefined views to subsets of the weaknesses archived. In the information retrieval context the relevant view is the one pointing to the CWEs used by NVD. Due to recent changes made~\cite{MITRE18b}, however, the predefined NVD-specific view is not suitable. Therefore, data is used from the CWE database only for the $102$ weaknesses that are present also in NVD indirectly via CVE mappings. Although the present context is weaknesses, the idea here is similar to the enforcement of ``one-to-one vulnerabilities'' between vulnerability databases~\cite{WenZhang15}. In contrast to some previous studies \cite{HanLi18}, all textual information is used for constructing the corpora. This information contains also meta-data strings, but the results reported did not differ much from those obtained by including only fields specific to natural language. The pre-processing described later on also filters out much of the meta-data.

The third and final source is the so-called Snyk database used for tracking security issues in open source software packages particularly in the web development context~\cite{Snyk18}. In contrast to the primary package managers used in Linux distributions, Snyk targets the secondary package managers and their repositories that are specific to programming languages. Although the Snyk database has seldom been used for research purposes, the underlying repositories have been studied extensively (see \cite{Mitropoulos14}, \cite{Raemaekers17}, and \cite{Squire18}, for instance). The following four repositories are included in the dataset assembled: \textit{Maven} (Java), \textit{pip} (Python), \textit{npm} (JavaScript), and \textit{RubyGems} (Ruby). In addition to the web development context, these repositories were selected due to sufficient amounts of vulnerabilities reported for the packages within the repositories. Therefore, the selection used allows to check whether the results are specific only to some repositories.

\begin{figure}[th!b]
\centering
\includegraphics[width=8.5cm, height=3.5cm]{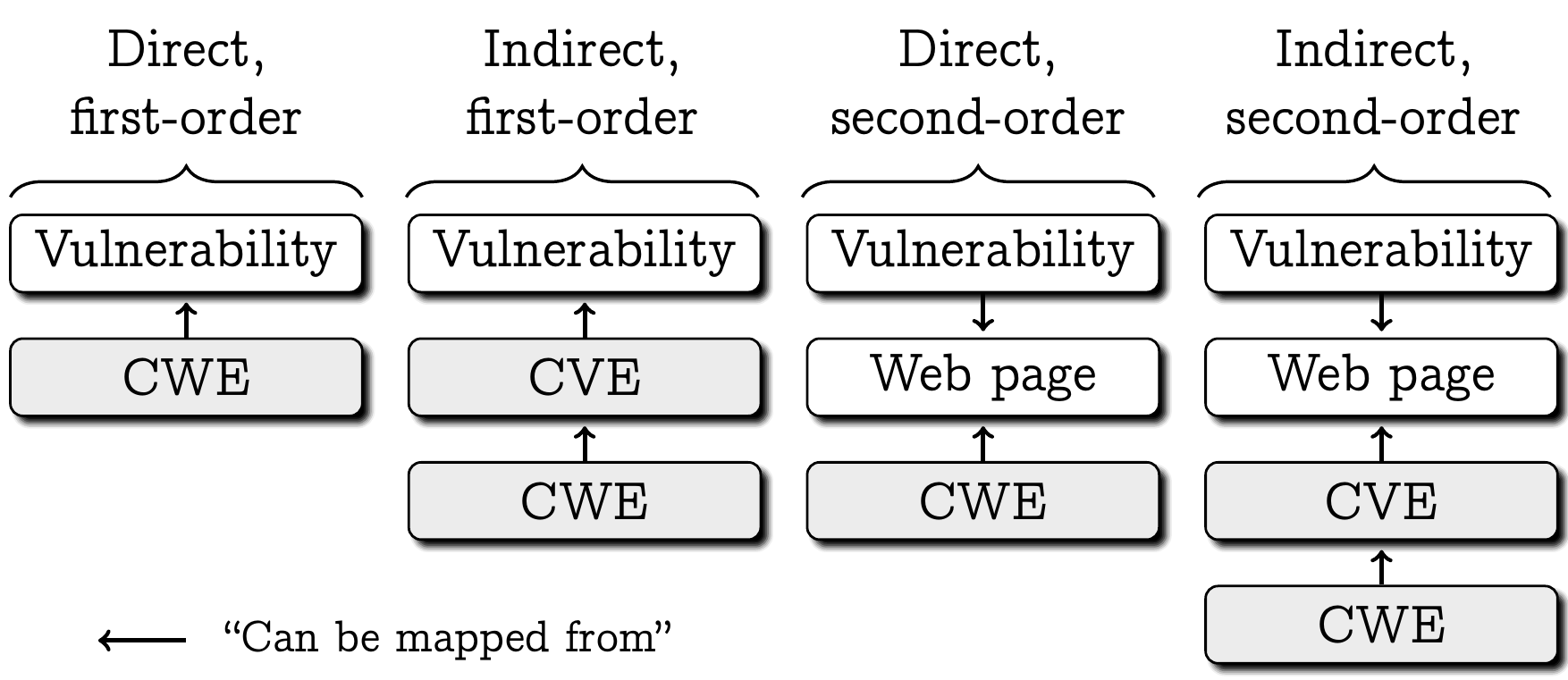}
\caption{An Example of Four Abstract Relations for Software Weaknesses}
\label{fig: relations}
\end{figure}

As is typical in vulnerability tracking \cite{DuRen18, RomanMunoz18, Ruohonen17TIR}, the Snyk database contains first-order and (online) second-order relations. As illustrated in Fig.~\ref{fig: relations}, these relations can be either direct or indirect with respect to CWE and CVE identifiers. For instance, the following vulnerability report (pymongo/40183) in the Snyk database represents a second-order indirect relation because the CWE in question can be mapped from the CVE visible in the link pointing to NVD:

\begin{small}
\begin{verbatim}
## Overview
[`pymongo`](https://pypi.python.org/pypi/pymongo) is a Python driver for
MongoDB.

`bson/_cbsonmodule.c` in the mongo-python-driver (aka.  pymongo) before
2.5.2, as used in MongoDB, allows context-dependent attackers to cause a
denial of service (NULL pointer dereference and crash) via vectors related
to decoding of an "invalid DBRef."

## References
- [NVD](https://web.nvd.nist.gov/view/vuln/detail?vulnId=CVE-2013-2132)
- [Github Commit](https://github.com/mongodb/mongo-python-driver/commit/
a060c15ef87e0f0e72974c7c0e57fe811bbd06a2)
\end{verbatim}
\end{small}

The first-order relations refer to the primary textual content stored to the database. As can be seen from the above excerpt, Snyk's maintainers archive each vulnerability with a brief textual description, which can be potentially mapped to a CWE with different text mining techniques. When constructing the corpora, all textual information is used except comment fields (lines starting with a \#) and links to further online material (lines starting with a dash). The additional online material constitute the second-order relations. Most vulnerabilities archived to Snyk are accompanied with one or more links pointing to security advisories, blogs, mailing lists, bug trackers, version control systems, hosting services such as GitHub, and other vulnerability databases, including NVD in particular. The content behind each of these links was downloaded, and for each successful download, all textual information was further added to the corpora sans the hypertext markup language elements. Given the terseness of the primary (first-order) textual information in the Snyk database, the additional (second-order) online material is beneficial for enlarging the corpora observed.

Thus, to summarize, the dataset contains textual information about software weaknesses from the CWE database, and first-order textual information about vulnerabilities from the Snyk database, as well as second-order textual information from online sources referenced in the Snyk database. The goal is to validate how well the vulnerabilities can be mapped to the CWEs. For this validation task, empirical estimation is carried out in a specific subset of the dataset.

\subsection{Estimation Subset}

A basic difficulty in classifying text documents is the typical absence of a ground truth against which text mining results can be compared. A typical way to tackle the issue is to compare text mining results against labels made by human experts~\cite{dosSantos15}. The same applies in the vulnerability context~\cite{DuRen18, Peclat18}. However, the approach adopted takes a different path: the text mining results are compared against simple but relatively robust regular expression searches. For each vulnerability, the following two simple regular expressions were used to search for direct CWE mappings and indirect ``CWE-from-CVE-in-NVD'' mappings:
\begin{equation}
\verb! (?:CWE)[-][0-9]{1,} !
~~\textmd{and}~~
\verb! (?:CVE|CAN)[-][0-9]{4}-[0-9]{4,} ! .
\end{equation}

%\begin{verbatim}
%(?:CWE)[-][0-9]{1,}
%\end{verbatim}

Only if a vulnerability matched only once from either one of the expressions, it was qualified to the estimation subset alongside the corresponding CWE, provided that the CWE was present among the $102$ weaknesses in NVD. Thus, each vulnerability in the estimation subset can be mapped with regular expression searches to a single unique weakness. It can be further noted that techniques such as majority-voting are unsuitable. The reason is simple. Many of the second-order relations point to web pages that catalog all CVEs assigned for a given package. By per-vulnerability voting based on the frequency of CVEs (or CWEs) mentioned in such pages, the uniqueness condition would be lost. Such voting would presumably also lead to haphazard mappings. Given these remarks,
\begin{equation}\label{eq: n}
n_1 = 82
\textmd{~weaknesses~and~}
n_2 = 585
\textmd{~vulnerabilities}
\end{equation}
were qualified to the estimation subset. For comparing the regular expression searches against information retrieval techniques, a common metric can be used:
\begin{equation}\label{eq: precision}
\textmd{Precision} = 
\frac{(\#~\textmd{same CWE})}{(\#~\textmd{same CWE}) +
(\#~\textmd{different CWE})} ,
\end{equation}
where the numerator denotes the frequency of vulnerabilities that were assigned to the same CWEs by both the regular expression and information retrieval techniques. The second term in the denominator refers to the frequency of vulnerabilities assigned to different CWE identifiers by the two techniques. Due to the described operationalization of the estimation subset, additional metrics (such as recall and accuracy) are not meaningful---it cannot be deduced whether the remaining vulnerabilities in the Snyk database are relevant or irrelevant with respect to CWEs. That said, it would be possible to use more metrics by splitting the estimation subset further into training and test sets~\cite{Murtaza16}, but, as will be seen, the estimation subset and the precision metric are alone sufficient for summarizing the paper's empirical validation results.

An important further remark should be made. Although the two distinct terms in \eqref{eq: precision} connote with ``true positives'' and ``false positives'', these terms do not convey their usual meanings in the present context: it is impossible to say with certainty which one of the two techniques is (in)valid. On one hand, the whole vulnerability tracking infrastructure is based on the unique CVE identifiers that are easy to search based on keywords~\cite{Jiminez16, Ruohonen17IWSMMensura}. Therefore: if a vulnerability archived to Snyk can be mapped with a regular expression search to a unique CWE either directly or indirectly via a CVE identifier, there is a fairly good chance that the vulnerability truly reflects the given software weakness. This argument is reinforced by the sample: many of the vulnerabilities archived to Snyk are accompanied with CVE-specific links to NVD's website, which, in turn, contains also the CWEs assigned for the CVEs in question. On the other hand, some vulnerabilities archived to the Snyk database contain explicit notes that these particular vulnerabilities differ from those archived to NVD with some particular CVE identifiers. Such remarks cause some false positives for the regular expression searches but not for the information retrieval techniques.

\subsection{Pre-processing}\label{subsec: pre-processing}

The pre-processing routine is fairly typical. To begin with, all textual data was \textit{transformed} by removing hypertext markup language elements and then lowercasing all letters. After the transformation, the data was \textit{tokenized} according to white space, punctuation characters, and other elements separating word boundaries. The resulting tokens were subsequently \textit{trimmed} by excluding tokens containing non-alphabetic characters, tokens that are common stop words, as well as tokens with length less than three or more than twenty characters. The remaining tokens were finally \textit{stemmed} with the Porter's~\cite{Porter80} classical algorithm. The NLTK library \cite{NLTK18} was used for the tokenization, stop words, and stemming. 

The stemmed tokens were used for three types of ``bag-of-words'' matrices:
\begin{equation}\label{eq: weight matrix}
\mtw_{(k)} ,
\quad \textmd{where}\quad 
k \in \lbrace 1, 2, 3 \rbrace .
\end{equation}

An element $w_{(k)ij}$ in a $\mtw_{(k)}$ denotes the \textit{weight} given for the $j$:th tokenization-based \textit{$k$-gram} (i.e., a unigram, a bigram, or a trigram) in the $i$:th \textit{document} (i.e., either a given CWE or a given vulnerability). The number of rows remains constant in each matrix: $i = 1, \ldots, n_1 + n_2$, where $n_1$ and $n_2$ are defined in \eqref{eq: n}. Given that the use of bigrams and trigrams substantially enlarges the amount of data, the number columns varies: $j = 1, \ldots, m_k$, where $m_k$ denotes the number of unique $k$-grams observed. It should be also noted that before assigning the actual weights, the underlying abstract data structures were pruned by excluding those $k$-grams that were present in a given corpus only twice or less. %If $\tilde{\mtw}_{(k)}$ denotes a raw weight matrix before pruning, this exclusion amounts to deleting those columns with a column sum less than three, and subsequently deleting the rows with a row sum of zero. 

\subsection{Weights}

Five different weights are used for the empirical validation. The \textit{term} frequency (that is, in the present context, $k$-gram frequency) provides the starting point:
\begin{equation}\label{eq: tf}
\TF : w_{(k)ij} = f_{(k)ij},
\quad w_{(k)ij} \in \mtw_{(k)},
\end{equation} 
where $f_{(k)ij}$ denotes the number of occurrences of the $j$:th $k$-gram in the document $i$. In addition, two simple TF-derivatives are empirically explored:
\begin{align}
\TFLOG : w_{(k)ij} &= \log(f_{(k)ij} + 1)
~~~\textmd{and}
\\
\TFBOOL : w_{(k)ij} &= 
\begin{cases}
0~~\textmd{if}~~f_{(k)ij} = 0 , \\
1~~\textmd{if}~~f_{(k)ij} > 0 .
\end{cases}
\end{align} 

Although there is no particular prior reason to expect that the results would differ with respect to these three simple weighting schemes, the usual rationale for $\TFLOG$, for instance, is based on the reasoning that the importance of a term is unlikely to grow linearly as implied by the TF weights \cite{Paik13}. The same rationale works also behind the current \textit{de~facto} weighting scheme, the so-called term-frequency-inverse-document-frequency (TF-IDF). Despite of the scheme's complex name, the actual weighting is simple:
\begin{align}\label{eq: tf-idf}
\TFIDF :~& w_{(k)ij} = f_{(k)ij} \times \omega_{(k)j} , 
\quad\textmd{where}
\\ \notag
& \omega_{(k)j} = \log([n_1 + n_2 + 1]~/~[\tilde{n}_{(k)j} + 1]) + 1
\end{align}
and $\tilde{n}_{(k)j}$ denotes the number of documents containing the $j$:th term (that is, $k$-gram). The IDF term, $\omega_{(k)j}$, used in \eqref{eq: tf-idf} is one of the many smoothed variants of the standard IDF formula. With smoothing or no smoothing, the rationale behind the IDF term is to penalize the occurrence of common terms. Even though the \text{TF-IDF} weights perform well in many applied problems, the same rationale has been used to define also many other more or less analogous weights \cite{JinChai05, Paik13}. To empirically explore one of these IDF-based derivatives, the so-called (pivoted) document length normalization (DLM) is as a good choice as any. It is given by
\begin{equation}\label{eq: dlm-idf}
\DLMIDF : w_{(k)ij} =
\left[\frac{1 + \log(f_{(k)ij} + 1)}{(1 - \beta) + \beta(L_{(k)i} / \overline{L}_{(k)})}\right] \omega_{(k)j} ,
\end{equation}
where $\beta$ is fixed to a scalar $0.2$, $L_{(k)i}$ denotes the length of the $i$:th document (defined as the sum of all term frequencies), and $\overline{L}_{(k)}$ refers to the average document length in a whole corpus \cite{Ibrahim16}. The five different weights enumerated are used for disseminating the subsequently discussed empirical validation results.

\section{Results}\label{section: results}

The five different weights from \eqref{eq: tf}~to~\eqref{eq: dlm-idf} were assigned to each of the matrix types noted in \eqref{eq: weight matrix}. This assignment resulted in fifteen weight matrices that establish the basis for the empirical results. For computing the similarities between the CWEs and vulnerabilities observed, the standard cosine similarity is used for each of the matrices. If $\tilde{\mtc}$ denotes a document-by-document matrix of cosine similarities, a $n_2 \times n_1$ vulnerability-by-weakness matrix $\mtc$ can be obtained by deleting $n_1$ rows from $\tilde{\mtc}$ that refer to CWEs, and carrying out an analogous operation for the columns. Then, each vulnerability (row) is mapped to the corresponding maximum row value in $\mtc$. If ties are present (meaning that a vulnerability has the same maximum cosine similarity with two or more CWEs), the vulnerability is mapped to the CWE picked by the regular expression searches, provided that any of the maximum values map to this particular CWE identifier.

The so-called latent semantic analysis~(LSA) was also briefly examined as an additional validation check. Although the usefulness for applied problems has been debated \cite{Gamallo11}, LSA builds on the rationale that a rank-reduced similarity matrix based on linear combinations may better reveal the underlying latent structure of a corpus. Computation is simple but expensive: LSA is a straightforward application of the  fundamental singular value decomposition, $\tilde{\mtc} = \mtu\mts\mtv^T$. Without delving into the linear algebra details, the actual reduction is also simple: a predefined number of the descending singular values in the diagonal matrix $\mts$ is restricted to zero, after which $\tilde{\mtc}$ is reconstructed by using the manipulated diagonal matrix~\text{\cite{dosSantos15, Hussain15}}. A simple heuristic was used for the manipulation: all singular values less than or equal to one were set to zero. For each of the fifteen matrices, roughly about a quarter of the singular values satisfied this heuristic.

\begin{figure}[p!]
%\begin{figure}[th!b]
\centering
\includegraphics[width=12cm, height=9cm]{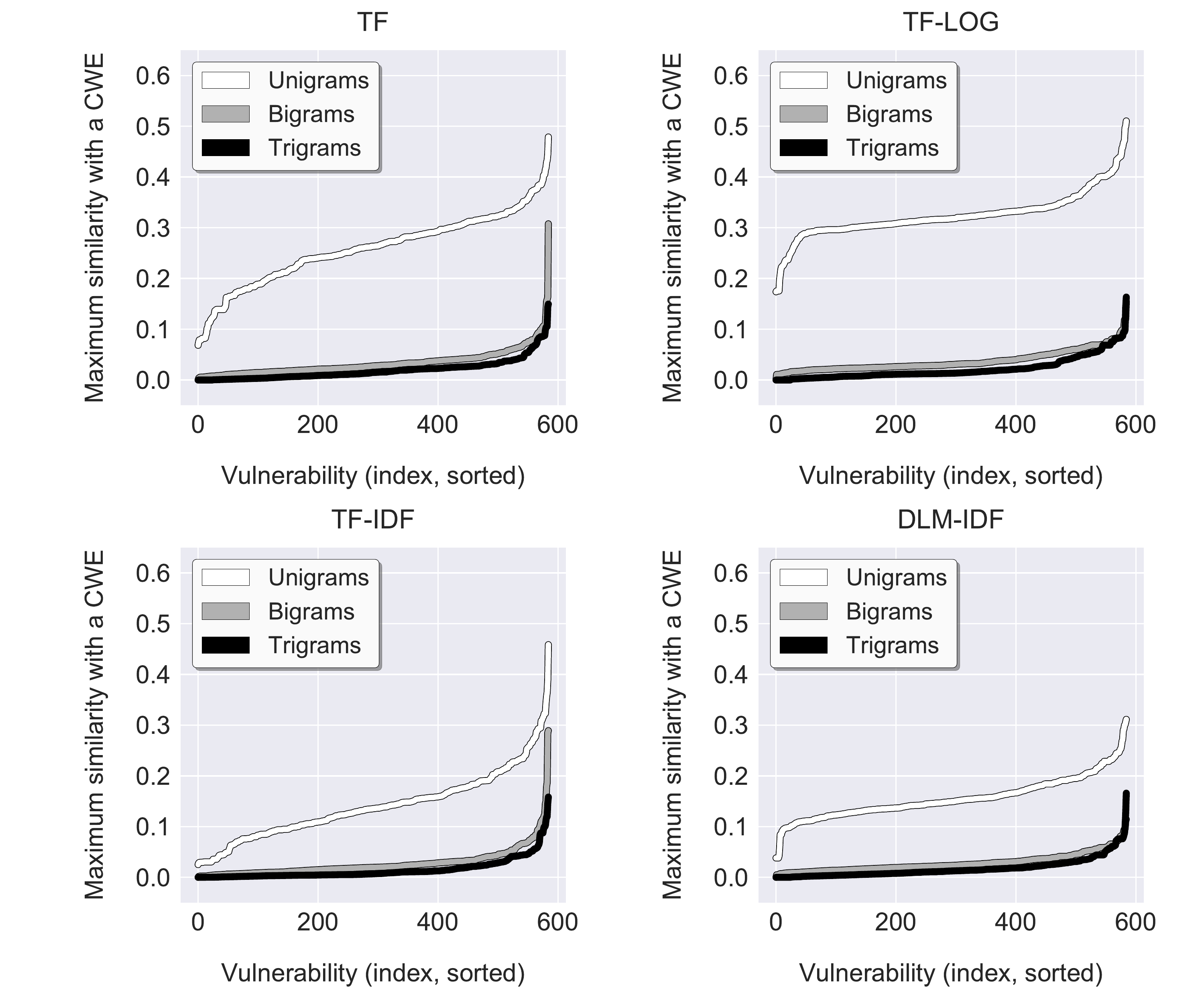}
\caption{Maximum Cosine Similarities According to Four Weights}
\label{fig: maxsim}
%\end{figure}
%
\vspace{15pt}
%
%\begin{figure}[th!b]
\centering
\includegraphics[width=11cm, height=8cm]{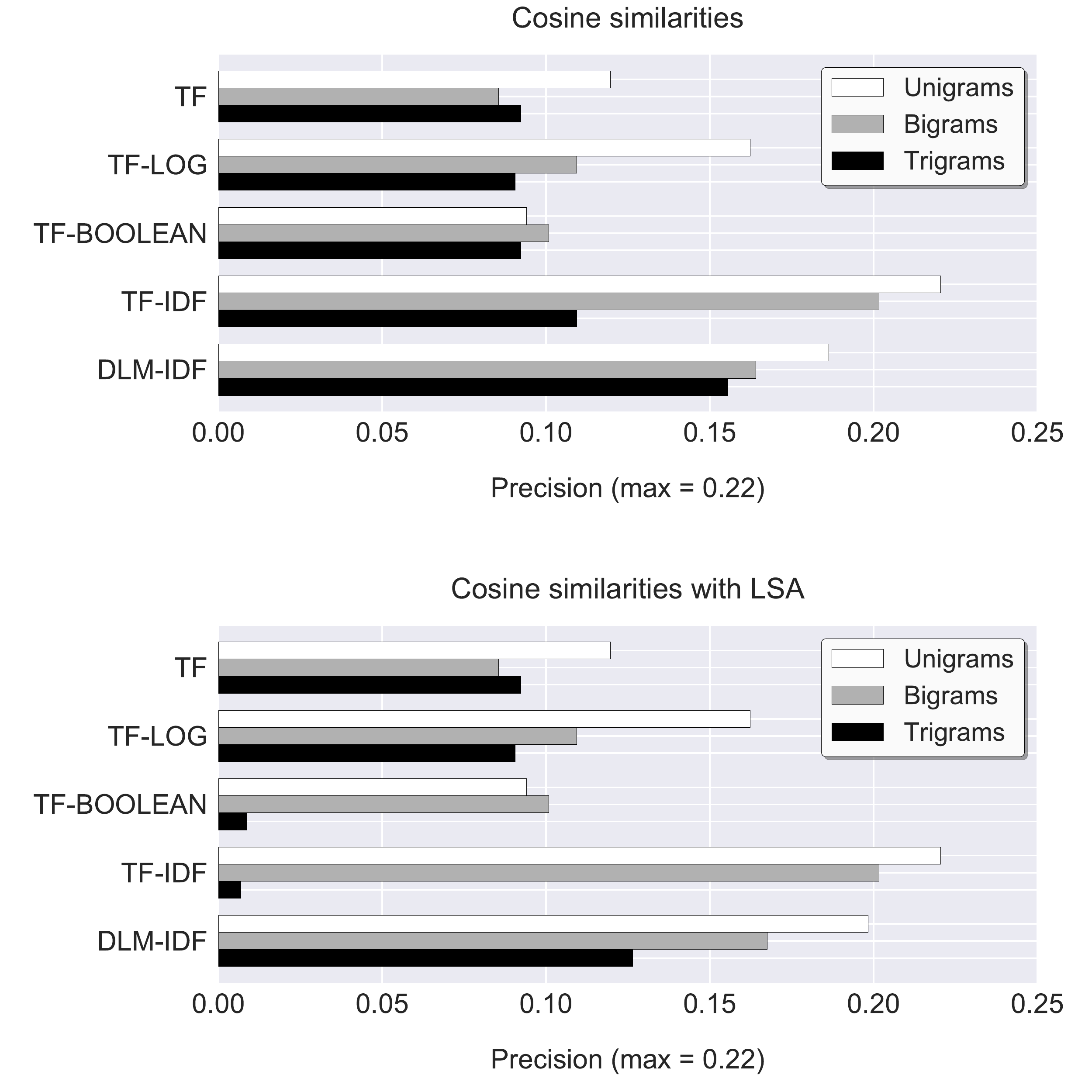}
\caption{Precision with Five Weights}
\label{fig: precision}
\end{figure}

Given these computational remarks, the results indicate only modest similarities between the CWEs and the vulnerabilities archived to Snyk. To illustrate this observation, Fig.~\ref{fig: maxsim} displays the maximum cosine similarities with four weights. Unigrams perform much better than bigrams or trigrams regardless of the weights used. The likely explanation relates to the numbers in Table~\ref{tab: descriptives}, which shows the number of columns in $\mtw_{(k)}$ and the average document lengths used for the DLM-IDF weights in~\eqref{eq: dlm-idf}. That is: even after the pruning of the weight matrices, many of the bigrams and trigrams appear only in few documents. The TF and \text{TF-LOG} weights also attain higher maximum similarities than the two IDF-based weights shown in the two plots on the bottom row in Fig.~\ref{fig: maxsim}. On average, however, the clear majority of the cosine similarities are well below $0.5$ for all weights used. These low similarity values translate into poor precision. As illustrated in Fig.~\ref{fig: precision}, the maximum precision is as low as $0.22$. The \text{TF-IDF} weights perform the best in this regard. LSA does not improve the precision.

\begin{table}[th!b]
\centering
\caption{Descriptive Statistics}
\label{tab: descriptives}
\begin{tabular}{lcrcrcr}
\toprule
&& Unigrams &\qquad\quad& Bigrams &\qquad\quad& Trigrams \\
\hline
Unique $k$-grams ($m_k$) 
&& 8435 && 32166 && 31745 \\
Average document length ($\overline{L}_{(k)}$) 
&& 1095 && 935 && 839 \\
\cmidrule{3-7}
~$\bullet$~Average CWE length 
&& 424 && 357 && 252 \\
~$\bullet$~Average vulnerability length 
&& 1175 && 1016 && 921 \\
\bottomrule
\end{tabular}
\end{table}

\begin{table}[th!b]
\centering
\caption{Average per-Repository Precision (TF-IDF)}
\label{tab: package precision}
\begin{tabular}{lccrcrcr}
\toprule
%& \multicolumn{7}{c}{Repository} \\
& \textit{Maven} && \textit{pip} && \textit{npm} &&  \textit{RubyGems} \\
\cmidrule{2-8}
Unigrams \qquad\qquad & $0.17$ &\quad\qquad& $0.34$ &\quad\qquad& $0.55$ &\quad\qquad& $0.25$ \\
Bigrams & $0.16$ && $0.31$ && $0.09$ && $0.62$ \\
Trigrams & $0.10$ && $0.12$ && $< 0.01$ && $0.50$ \\
\bottomrule
\end{tabular}
\end{table}

All in all, the validity of common textual information retrieval techniques seems questionable for software weaknesses. That said, the results diverge considerably between the four repositories (see Table~\ref{tab: package precision}). While the results are consistently poor particularly for \textit{Maven} and to a lesser extent \textit{pip}, an average precision of $0.55$ is obtained for \textit{npm}. Interestingly, the use of bigrams raises the average precision to $0.62$ for \textit{RubyGems}. Thus, it can be also concluded that the poor precision cannot be generalized to all repositories or programming languages.

\section{Discussion}\label{section: discussion}

This paper presented a preliminary validation of common textual information retrieval techniques for mapping vulnerabilities to weaknesses. When compared to basic regular expression searches, the techniques seem to perform poorly according to the results based on four repositories tracked in the Snyk vulnerability database.  For searching specific vulnerabilities or weaknesses from software repositories, simple keyword searches based on CVE and CWE identifiers seem more robust. These commonly used \cite{DuRen18, Jiminez16, Ruohonen17IWSMMensura} domain-specific searches could be augmented by the information retrieval techniques \cite{Fautsch10}, however. In other words: it might be possible to prefer the regular expression searches as a primary retrieval technique and use the information retrieval techniques as a secondary method for retrieving additional content not captured by the keyword-based searches. It is also important to stress that the results vary across repositories. This observation hints that the choice over particular security-related corpora has likely a strong effect upon the vulnerability-CWE mappings.

\begin{figure}[th!b]
\centering
\includegraphics[width=10cm, height=5.5cm]{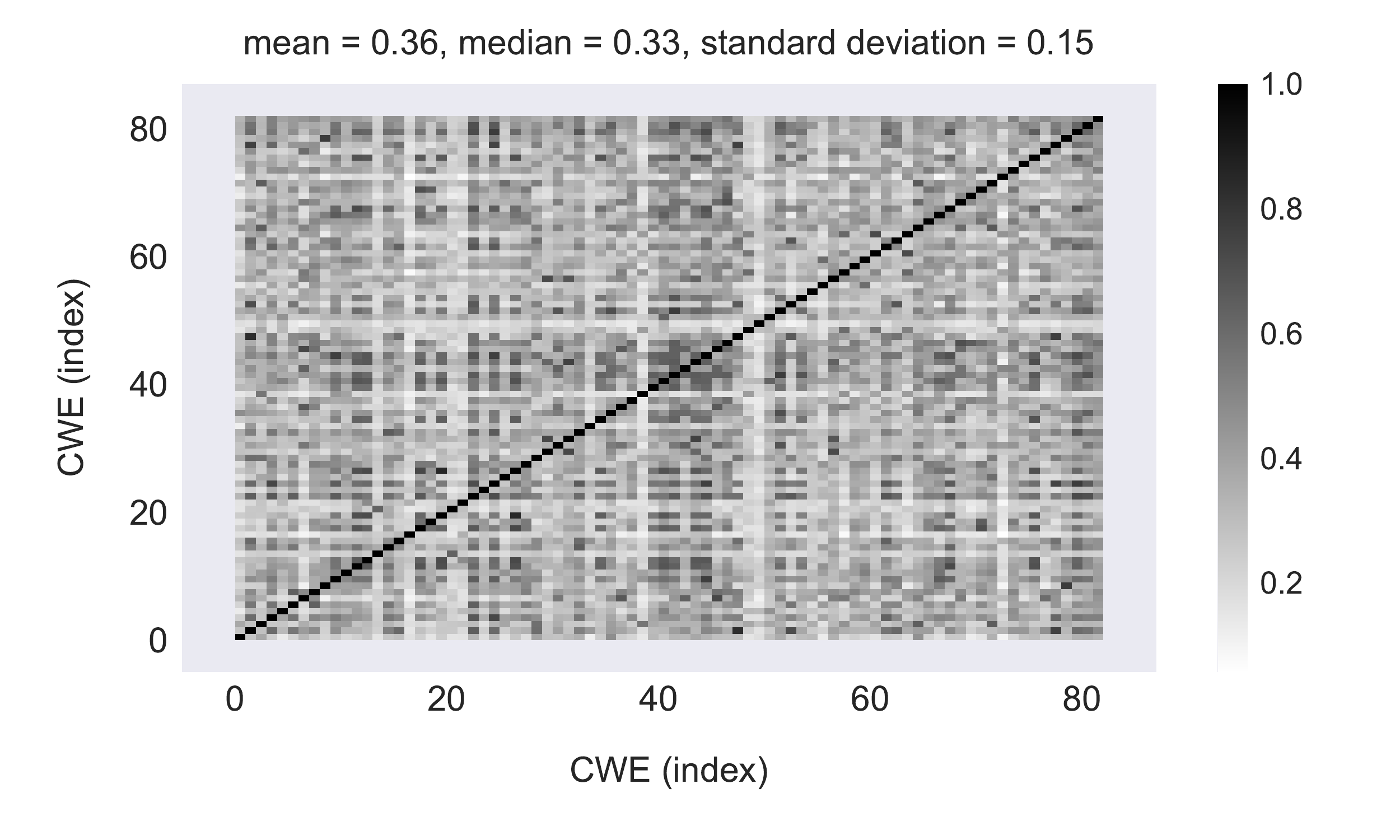}
\caption{Similarities Between Weaknesses (TF-IDF, unigrams)}
\label{fig: heat}
\end{figure}

Three points can be noted about limitations and directions for further research. First, as CWE is a hierarchical classification system, it might be argued that larger ontology-based weakness groups should be used. However, all of the individual CWEs observed in this paper have been tracked in NVD, which undermines the rationale for using such CWE groups in the present context. The construct validity is also undermined by the illustration in Fig.~\ref{fig: heat}, which shows that the CWEs observed are not very similar with respect to each other.

Second, many computational checks could be done to further validate the vulnerability-CWE mappings. For instance, the cosine similarity used could be verified against other commonly used similarity metrics. (It can be noted that the so-called Jaccard similarity performs even worse, however.) Further examples include the calibration of the LSA's reduction step, the use of a more aggressive stemming algorithm, and the examination of part-of-speech tagging.  

Third, it seems reasonable to try to either enlarge or enrich the datasets used for validation. The textual information stored to Snyk, CWE, OWASP, and related databases is terse in terms of natural language and prose, but there are technical terms that should be specifically weighted. If the small excerpt shown in Subsection~\ref{subsec: data sources} is taken as an example, the trigrams \texttt{denial of service} and \texttt{NULL pointer dereference} should attain higher weights than any of the other $k$-grams. Such domain-specific weights entail the construction of a reference corpus. Given the generally poor precision reported and the variance across repositories, it may also be that neither CWE nor OWASP are ideal for a construction of a reference corpus. Instead,  language-specific guides for secure programming~\cite{CERT18a} may be more suitable, among other potential sources related to software weaknesses and vulnerabilities. As an alternative to enrichment, big data analysis is also plausible. Due to the central role of CVE identifiers (cf.~Fig.~\ref{fig: relations}), web crawling could be used to gather a truly massive dataset for text mining. Recent work~\cite{DuRen18} shows also some promise for web crawling approaches. But the larger the datasets, the coarser the mappings, and the bigger the validity concerns.

\bibliographystyle{splncs03}
%\bibliography{vuln}

\end{document}